\documentstyle[12pt]{article}
\begin{document}
\newcommand{\bea}{\begin{eqnarray}}
\newcommand{\eea}{\end{eqnarray}}
\newcommand{\be}{\begin{equation}}
\newcommand{\ee}{\end{equation}}
\newcommand{\non}{\nonumber}
\newcommand{\ov}{\overline}
\global\parskip 6pt
\begin{titlepage}
\flushright{DIAS-STP-95-28/July 1995}\\
\begin{center}
{\Large\bf On the Quantum Kinetic Equation}\\
\vskip .10in
{\Large\bf in Weak Turbulence}\\
\vskip 0.5in
Mark Rakowski \footnote{Email: Rakowski@maths.tcd.ie} \\
\vskip .10in
{\em Dublin Institute for Advanced Studies, 10 Burlington Road, }\\
{\em Dublin 4, Ireland}\\
\vskip .10in
and \\
\vskip .10in
Siddhartha Sen \\
\vskip .10in
{\em School of Mathematics, Trinity College, Dublin 2, Ireland} \\
\end{center}
\vskip .10in
\begin{abstract}
  The quantum kinetic equation used in the study of weak turbulence
is reconsidered in the context of a theory with a generic quartic
interaction. The expectation value of the time derivative of the
mode number operators is computed in a perturbation expansion which
places the large diagonal component of the quartic term in the
unperturbed Hamiltonian. Although one is not perturbing around a free
field theory, the calculation is easily tractable owing to the fact
that the unperturbed Hamiltonian can be written solely in terms of
the mode number operators.
\end{abstract}
\vskip 1.5in
\begin{center}
\end{center}

\end{titlepage}

\section{Introduction}

   In one approach to the statistical description of weak turbulence,
a central role is played by the kinetic wave equation \cite{Zak,Dya}.
This equation for the time derivative of the mode numbers has been derived
for both classical and quantum systems in a perturbation series by
expanding about a free field (harmonic oscillator) theory. In this
paper, we will reconsider this derivation for a quantum mechanical
system whose Hamiltonian is a sum of generic quadratic and quartic
terms. Our perturbation expansion will perturb around an operator
which contains the diagonal component of the quartic term together
with the usual quadratic, or free field, component. Since one is
interested in the expectation values of fields between states with
large mode numbers, it is sensible to include as much of these in the
unperturbed Hamiltonian as the calculation permits. We do not need
to assume that the coupling to all the quartic terms in the Hamiltonian
is small; the diagonal part can be arbitrarily large in our approach.

   We begin in the following section with a precise statement of the
theory under consideration and an encapsulation of the interaction
picture used to carry out the derivation of the quantum kinetic equation.
At this order, the largest terms are cubic in the mode numbers and these
we calculate explicitly. We will find some additional terms not present
in another derivation of the quantum kinetic equation \cite{Zak}, and
then go on to consider under what conditions one might expect those
corrections to be small. A discussion of
stationary solutions to the kinetic equation then follows.

\section{The Quantum Kinetic Equation}

Let us consider a quantum mechanical system based on the Hamiltonian,
\bea
H= \sum_{k}\; \omega_{k}\, a^{\dagger}_{k}\, a_{k}    +
   \sum_{k_{1},\cdots,k_{4}} \; T_{k_{1}k_{2}k_{3}k_{4}} \;
   a^{\dagger}_{k_{1}}\, a^{\dagger}_{k_{2}}\, a_{k_{3}}\, a_{k_{4}}
   \;\; , \label{H}
\eea
which contains generic quadratic and quartic terms; the number $d$ of
spatial dimensions in which the system evolves is arbitrary. The free field
oscillator energies $\omega_{k}$ are assumed given and constitute
part of the specification of the system. The function
$T_{k_{1}k_{2}k_{3}k_{4}}$ includes, by definition, the momentum
conserving factor $\delta_{k_{1}+k_{2},k_{3}+k_{4}}$
(both the vector $k_{i}$ and the $\delta$-function should be understood
as $d$-dimensional quantities). Beyond the implicit symmetry properties
which are that $T_{k_{1}k_{2}k_{3}k_{4}}$ is symmetric in the first
two and last two indices, and that under complex conjugation
$T^{*}_{k_{1}k_{2}k_{3}k_{4}} = T_{k_{3}k_{4}k_{1}k_{2}}$, this
coefficient
may contain further momentum dependence which we will otherwise not
restrict in the derivation of the kinetic equation. As usual,
$a^{\dagger}_{k}$ and $a_{k}$ in eqn. (\ref{H}) denote Bose creation
and annihilation operators, and they obey the commutation relation,
\bea
[ a_{k}, a^{\dagger}_{l} ] = \delta_{k,l} \;\; .
\eea

It is
convenient to use the number operator $\hat{n}_{k} = a^{\dagger}_{k}\,
a_{k}$ for each mode in our system. The states which diagonalize these
number operators satisfy \cite{FW}:
$\hat{n}_{k} | n_{k} \rangle = n_{k} \, | n_{k} \rangle$,
$a_{k}\, | n_{k} \rangle = \sqrt{n_{k}}\, | n_{k} -1 \rangle$, and
$a^{\dagger}_{k}\, | n_{k} \rangle = \sqrt{n_{k}+1}\, | n_{k}+1 \rangle$,
and are clearly labelled by their eigenvalues.

A perturbation series (see \cite{FW} for a thorough exposition) begins
by splitting the Hamiltonian into an unperturbed component $H_{0}$ and
a ``small'' component $H_{1}$. In our approach, we will place the
diagonal of the quartic component in the unperturbed sector, so that
$H = H_{0} + H_{1}$ with,
\bea
H_{0} &=& \sum_{k}\; (\omega_{k}- 2\, T_{k}) \, \hat{n}_{k}    +
       2\; \sum_{k,l}\; T_{kl}\;
       \hat{n}_{k}\, \hat{n}_{l}  \\
H_{1} &=& \sum_{k_{1},\cdots,k_{4}}\;
   T^{\prime}_{k_{1}k_{2}k_{3}k_{4}} \;
   a^{\dagger}_{k_{1}}\, a^{\dagger}_{k_{2}}\, a_{k_{3}}\, a_{k_{4}}
\;\; , \non
\eea
and where we have introduced the notation $T_{k} = T_{kkkk}$,
$T_{kl} = T_{klkl}$, and
\bea
T^{\prime}_{k_{1}k_{2}k_{3}k_{4}} = \left\{ \begin{array}{ll}
T_{k_{1}k_{2}k_{3}k_{4}}     &\mbox{if $k_{1}\not= k_{3}$ or $k_{4}$} \\
0                            &\mbox{otherwise}
\end{array}
\right. \;\; .
\eea
We can express this equivalently as,
\bea
T^{\prime}_{k_{1}k_{2}k_{3}k_{4}} = (\;
1 - \delta_{k_{1}k_{3}}\, \delta_{k_{2}k_{4}} -
    \delta_{k_{1}k_{4}}\, \delta_{k_{2}k_{3}} +
    \delta_{k_{1}k_{2}}\, \delta_{k_{1}k_{3}}\, \delta_{k_{2}k_{4}} \; )
\; T_{k_{1}k_{2}k_{3}k_{4}}\;\; .
\eea
It is important to emphasize that we need not assume that the
coefficient functions $T_{kl}$ be small, as they are part
of the unperturbed Hamiltonian.
The validity of the perturbation expansion depends, however, that
$T^{\prime}_{k_{1}k_{2}k_{3}k_{4}}$ be ``small'' relative to the mode
numbers.

In the Heisenberg representation of quantum mechanics, operators are
time dependent while the states are time independent. Given some
operator $A$, its time evolution as a Heisenberg operator $A_{H}(t)$
satisfies
\bea
\frac{d}{dt} A_{H}(t) = i\; [ H, A_{H}(t) ]\;\; ,
\eea
and one may equivalently write this as
$A_{H}(t) = \exp[iHt]\; A_{H}(0)\; \exp[-iHt]$.
Any expectation value in this representation therefore satisfies,
\bea
\frac{d}{dt} \; \langle \Psi_{1} | A_{H}(t) | \Psi_{2} \rangle =
\langle \Psi_{1} |\;
\frac{d}{dt} A_{H}(t)\; | \Psi_{2} \rangle =
\langle \Psi_{1} |\,  i [ H, A_{H}(t) ] \, | \Psi_{2} \rangle \;\; ,
\eea
since the
states $| \Psi_{a} \rangle$ are independent of $t$.
We are interested in the case $A = \hat{n}_{k}$ for large $t$,
and we will compute,
\bea
\lim_{t\rightarrow \infty}\; \langle \Psi |\; i [ H, \hat{n}_{kH}(t) ]
\; |\Psi\rangle  \;\; ,                  \label{exval}
\eea
for some state $|\Psi\rangle$ which we
will later specify. This is the precise meaning we ascribe to
the time derivative of the mode number appearing in other presentations
\cite{Zak,Dya} of the kinetic wave equation.

For the purposes of perturbation theory, one moves to the interaction
picture where the following relations hold,
\bea
\langle  \Psi_{1} |\; {\cal O}_{H}(t) \;| \Psi_{2}  \rangle &=&
\langle \Psi_{1}(t) |\; {\cal O}_{I}(t) \;
| \Psi_{2}(t)  \rangle \label{Int}\\
| \Psi_{a}(t) \rangle &=& \exp[i\,H_{0}\, t ]\;
\exp[-i\,H\,(t-t')]\;\exp[- i\,H_{0}\, t']\; | \Psi_{a}(t') \rangle \non \\
& &\non \\
{\cal O}_{I}(t) &=& \exp[i\, H_{0}\, (t-t')] \; {\cal O}_{I}(t')\;
\exp[-i\, H_{0}\,(t-t')] \;\; . \non
\eea
for all operators ${\cal O}$ and all states $|\Psi_{a} \rangle$.
Interestingly, the time evolution of the operators
$a_{kI}^{\dagger}$ and $a_{kI}$
in this model is very simple in spite of the fact that
they do not evolve via a free field Hamiltonian. It is not difficult
to first show that
\bea
[ a_{k} , H_{0} ] = (\omega_{k} + 4 \sum_{l}\; T_{kl}\; \hat{n}_{l} )\;
a_{k}\;\; ,
\eea
and using this one can quickly prove,
\bea
a_{kI}(t) &=& \exp[i\, H_{0}\, t] \; a_{kI}(0)\;
\exp[-i\, H_{0}\,t] \\
& & \non \\
&=& \exp[-i\, t\,
(\omega_{k} + 4 \, \sum_{l} \, T_{kl}\, \hat{n}_{l} )]
\; a_{kI}(0) \non\\
&=& a_{kI}(0) \; \exp[-i\, t\, (\omega_{k} - 4\, T_{k}
+ 4 \, \sum_{l} \,
T_{kl}\, \hat{n}_{l} )]  \;\; . \non
\eea

   The combination of operators in the second line of (\ref{Int}) is
conveniently denoted by,
\bea
U(t,t') &=& \exp[i\,H_{0}\, t]\; \exp[-i\,H\,(t-t') ] \;
            \exp[- i\,H_{0}\, t'] \label{U} \\
&=& 1 - i\; \int_{t'}^{t}\; d\tau H_{1I}(\tau)\; U(\tau, t') \;\; .\non
\eea
To lowest order in the interaction $H_{1}$, we just set $U(\tau,t')=1$
on the right hand side of eqn. (\ref{U}). Our goal is to compute,
\bea
\langle \frac{d}{dt} \hat{n}_{k} \rangle \equiv
\lim_{t\rightarrow \infty}\; \langle \Psi(-t)|\; U^{\dagger}(t,-t)\;
{\cal O}_{I}(t)\;U(t,-t)\; |\Psi(-t)\rangle \;\; ,\label{exval2}
\eea
where ${\cal O} = i [ H_{1}, \hat{n}_{k} ]$ . We should emphasize that
the states on both sides of this expectation value are {\em in}-states
in the sense of scattering theory as both are at $-\infty$.

Noticing that $H_{0}$
commutes with $\hat{n}_{k}$, the computation of eqn. (\ref{exval2}) to
the lowest order in perturbation theory reduces to,
\bea
\lim_{t\rightarrow \infty}\; \langle \Psi(-t)|\;\;i[H_{1I}(t),
\hat{n}_{k}] + \int_{-t}^{t}\;[ [ H_{1I}(t), \hat{n}_{k} ] ,
H_{1I}(\tau) ] \;d\tau\;\; |\Psi(-t)\rangle \;\; .
\label{exval3}
\eea
We assume that the state $|\Psi(-\infty)\rangle$ is an eigenstate
of the number operators $n_{k}$. It is not difficult to show that
the first order term in (\ref{exval3}) does not contribute;
to see this one simply computes,
\bea
[H_{1}, \hat{n}_{k}] = 2\; \sum_{k_{1}k_{2}k_{3}}\; (
T^{\prime}_{k_{1}k_{2}k_{3}k}\;
a^{\dagger}_{k_{1}}\, a^{\dagger}_{k_{2}}\, a_{k_{3}}\, a_{k} -
T^{\prime}_{k_{1}kk_{2}k_{3}}\;
a^{\dagger}_{k_{1}}\, a^{\dagger}_{k}\, a_{k_{2}}\, a_{k_{3}}  )\;\; .
\label{Hn}
\eea
When we take this between the same states, there is no way to
pair the creation and annihilation operators as $T^{\prime}$ is
off-diagonal, and hence the first term in
(\ref{exval3}) manifestly vanishes.
However, we should point out that this term would also vanish even
if we had not subtracted out the diagonal, and had instead made the normal
perturbative expansion around the quadratic term in $H$; the cancellation
would then involve a mixing of both terms in eqn. (\ref{Hn}).

One of the ingredients needed in this computation is the expectation
value,
\bea
\langle a^{\dagger}_{k_{1}}\, a^{\dagger}_{k_{2}}
\, a_{k_{3}}\, a_{k_{4}}\,
a^{\dagger}_{l_{1}}\, a^{\dagger}_{l_{2}}\, a_{l_{3}}
\, a_{l_{4}} \rangle
\;\; ,
\eea
where $k_{1}$ or $k_{2}$ $\not=$ $k_{3}$ or $k_{4}$,
and $l_{1}$ or $l_{2}$ $\not=$ $l_{3}$ or $l_{4}$.
The states on both sides are identical and are eigenstates of
all the number operators. A simple computation yields,
\bea
(n_{l_{1}}+1)\; n_{l_{3}}
&[& \; ( \delta_{l_{1}k_{3}}\, \delta_{l_{2}k_{4}} +
           \delta_{l_{1}k_{4}}\, \delta_{l_{2}k_{3}} ) \;
          \; (n_{l_{2}} + 1) -
\delta_{l_{1}l_{2}}\,\delta_{l_{1}k_{3}}\,
\delta_{l_{2}k_{4}} \; n_{l_{2}}\; ] \cdot \non \\
&[& ( \delta_{l_{3}k_{1}}\, \delta_{l_{4}k_{2}} +
      \delta_{l_{3}k_{2}}\, \delta_{l_{4}k_{1}} )
\; n_{l_{4}} - \delta_{l_{3}l_{4}}\,
\delta_{l_{3}k_{1}}\, \delta_{l_{4}k_{2}}\;
(n_{l_{4}} + 1) \; ] \;\; . \label{nnnn} \;\;\;\;\;\;
\eea
We should note that it is not sufficient for the purposes of our
calculation, even working in the large $n_{l}$ limit, to keep only the
most dominant terms in the above expression which are fourth order
in these mode numbers. When this expression is used in our calculation,
we will see that the fourth order terms cancel and the next to leading
terms remain. For this reason we have been careful to take account
of the possibility that $l_{1}$ equals $l_{2}$, and so on, in this
expectation value; no assumptions (e.g. random phase approximation;
see \cite{Zak}) have been made in obtaining the expression in (\ref{nnnn}).

It is straightforward, though rather tedious, to assemble the
above pieces, and the tree level expression for the quantity
in (\ref{exval2}) is found to be,
\bea
\langle \frac{d}{dt} \hat{n}_{k} \rangle
&=& 8\,\pi \sum_{k_{1},k_{2},k_{3}}
| T^{\prime}_{k_{1}k_{2}k_{3}k}|^{2}\;
\delta(k_{1},k_{2},k_{3},k) \; [\; s_{3}(k_{1},k_{2},k_{3},k) \label{ndt}\\
& & \;\;\;\;\;\;\;\;\;\;\;\;\;\;\;\;\;\;\;\;\;\;\;
+ s_{2}(k_{1},k_{2},k_{3},k) + s_{1}(k_{1},k_{2},k_{3},k)\;  ]\;\; ,
\non
\eea
with the functions $s_{a}(k_{1},k_{2},k_{3},k)$ given by,
\bea
s_{3} &=&
4 \; ( \; n_{k_{1}}\, n_{k_{2}}\, n_{k_{3}}
+ n_{k_{1}}\, n_{k_{2}}\, n_{k}
- n_{k_{1}}\, n_{k_{3}}\, n_{k}
- n_{k_{2}}\, n_{k_{3}}\, n_{k} \; ) \\
&-& 2\; \delta_{k_{1}k_{2}} \;
( \; n_{k_{1}}\, n_{k_{2}}\, n_{k_{3}}
+ n_{k_{1}}\, n_{k_{2}}\, n_{k} \; )
  + 2\; \delta_{k_{3}k} \;
( \;  n_{k_{1}}\, n_{k_{3}}\, n_{k}
+ n_{k_{2}}\, n_{k_{3}}\, n_{k} \; ) \non\\
& & \non \\
s_{2} &=&
4 \; ( \; n_{k_{1}}\, n_{k_{2}} - n_{k_{3}}\, n_{k}\; )
 - 2\; \delta_{k_{1}k_{2}} \;
( \; n_{k_{1}}\, n_{k_{2}} + n_{k_{1}}\, n_{k} +
n_{k_{1}}\, n_{k_{3}} \; ) \non \\
&+& 2\; \delta_{k_{3}k} \;
( \; n_{k_{1}}\, n_{k_{3}} + n_{k_{3}}\, n_{k}
+ n_{k_{2}}\, n_{k_{3}}\; )   \non \\
& & \non \\
s_{1} &=& -2\; \delta_{k_{1}k_{2}} \; n_{k_{1}}
+ 2\; \delta_{k_{3}k} \; n_{k_{3}} \;\; .  \non
\eea
For large values of the mode numbers, the $s_{3}$ term which is
cubic in those variables will dominate.
The energy conserving delta function $\delta(k_{1},k_{2},k_{3},k)$,
which arose from an integration over $\tau$, is given by,
\bea
& & \delta(\; \omega_{k_{1}}+\omega_{k_{2}}-\omega_{k_{3}}-\omega_{k} +
4\; \sum_{l}\; ( T_{k_{1}l} + T_{k_{2}l} - T_{k_{3}l} - T_{kl} )\; n_{l}
\\
&+& 4\; ( T_{k_{3}} + T_{k} + T_{k_{1}k_{2}} + T_{k_{3}k}
- T_{k_{1}k} - T_{k_{2}k} - T_{k_{1}k_{3}} - T_{k_{2}k_{3}} ) \; )
\;\; . \non
\eea
It is obtained by moving the $\tau$ operator dependence
entirely to either the left or the right where it becomes a normal
function of $\tau$ after acting on the states, and then using the
representation,
$\int_{-\infty}^{+\infty}\; e^{i \tau x} \; d\tau = 2\pi\; \delta(x)$.
This rather asymmetric looking expression can be recast as,
\bea
&\phantom{+}& \delta(\; (\omega_{k_{1}} - 2T_{k_{1}}) +
(\omega_{k_{2}} - 2T_{k_{2}})
- (\omega_{k_{3}} - 2T_{k_{3}}) - (\omega_{k} - 2T_{k}) \\
&+& 2 \sum_{l}\; ( T_{k_{1}l} + T_{k_{2}l} - T_{k_{3}l} - T_{kl} )
(2 \, n_{l} + \delta_{k_{1}l} + \delta_{k_{2}l} - \delta_{k_{3}l}
- \delta_{kl} ) \; ) \;\; . \non
\eea
If we define an effective energy per state by
\bea
\varepsilon_{m} \;\; =\;\;  \omega_{m} - 2\, T_{m} + 2\; \sum_{l}\; T_{ml}\;
(\; 2\, n_{l} + \delta_{k_{1}l} + \delta_{k_{2}l} - \delta_{k_{3}l}
-\delta_{kl} \; )\;\; ,
\eea
then the delta function can be written most simply as,
\bea
\delta(\; \varepsilon_{k_{1}} + \varepsilon_{k_{2}} -
\varepsilon_{k_{3}} - \varepsilon_{k} \; ) \;\; .
\eea
The exact expression for $\varepsilon_{m}$ given above is greatly
simplified in the large $n_{l}$ situation in which we work, so that
\bea
\varepsilon_{m} \;\; \approx \;\;  \omega_{m} + 4\; \sum_{l}\;
T_{ml}\; n_{l} \;\; .  \label{eps}
\eea

   Let us emphasize that the leading order terms which are cubic in
the mode numbers in (\ref{ndt}) do not fully agree with the expressions
found in \cite{Zak,Dya}; our leading order terms can be written as,
\bea
\langle \frac{d}{dt} \hat{n}_{k} \rangle &=&
8\,\pi \sum_{k_{1},k_{2},k_{3}}
| T^{\prime}_{k_{1}k_{2}k_{3}k}|^{2}\;
\delta(\; \varepsilon_{k_{1}} + \varepsilon_{k_{2}} -
\varepsilon_{k_{3}} - \varepsilon_{k} \; )
\; \cdot \label{newnd} \\
& & \;\;\;\; n_{k_{1}}\, n_{k_{2}} \, n_{k_{3}} \, n_{k}
\; \{ \; 4\, ( \, \frac{1}{n_{k_{3}}} + \frac{1}{n_{k}} -
\frac{1}{n_{k_{1}}} - \frac{1}{n_{k_{2}}} \, ) \non \\
& & \;\;\;\;\;\;\;\; - 2\; \delta_{k_{1}k_{2}} \;
( \, \frac{1}{n_{k_{3}}} + \frac{1}{n_{k}} \, )
+ 2\; \delta_{k_{3}k} \;
( \,  \frac{1}{n_{k_{1}}} + \frac{1}{n_{k_{2}}} \, )\; \} \;\; ,  \non
\eea
with $\varepsilon_{m}$ given by eqn. (\ref{eps}).
One clear difference in our result concerns the last two terms in
(\ref{newnd}) which are not
present in \cite{Zak,Dya}; these would still be there in our analysis
even if we had perturbed around the quadratic term in $H$. In fact, further
diagonal terms would presumeably be present as well, but these we have
accounted for by incorporating them into $H_{0}$.
This discrepancy will
be discussed in the following section. The other clear difference is due to
our different perturbation expansion which results in the $\delta$-function
involving the effective energy $\varepsilon_{k}$, rather than $\omega_{k}$.

\section{Stationary Solutions}

  The analysis of stationary solutions to the kinetic equation in the usual
perturbative version can be found in \cite{Zak,Dya}. In that analysis, the
extra terms we found in (\ref{newnd}) are not considered, and they analyze
the condition,
\bea
0 &=& \sum_{k_{1},k_{2},k_{3}}
| T_{k_{1}k_{2}k_{3}k}|^{2}\;
\delta(\omega_{k_{1}} + \omega_{k_{2}} - \omega_{k_{3}} - \omega_{k} )
\cdot \\
& & \;\;\; n_{k_{1}}\, n_{k_{2}} \, n_{k_{3}} \, n_{k}
\; \{ \, \frac{1}{n_{k_{3}}} + \frac{1}{n_{k}} -
\frac{1}{n_{k_{1}}} - \frac{1}{n_{k_{2}}}  \; \} \;\; . \non
\eea
The question therefore arises as to whether the
other terms in eqn. (\ref{newnd}), which are also cubic
in the mode numbers, are small relative
to the others. In this regard, it is noteworthy that the additional terms
involve one less sum over momentum space. So if the terms in the summation
are large over some reasonable domain in momentum space, then those
factors will be suppressed by roughly the volume of that domain. Our
calculation thus far has assumed that at least some of the mode numbers
are large compared to unity and neglecting these additional two terms
amounts to some kind of additional condition such as the one just
suggested.
One might instead
entertain a more restricted class of $T^{\prime}_{k_{1}k_{2}k_{3}k_{4}}$
coefficient such that the additional terms vanish identically; perhaps
this might be natural in the context of vortex dynamics.
For example, it could contain the factor
$|k_{1}-k_{2}|^{\sigma}\cdot |k_{3}-k_{4}|^{\sigma}$,
for $\sigma > 0$. We will not
consider this issue further here, and will proceed to consider
the stationary solutions to eqn. (\ref{newnd}) assuming that the
last two terms can be neglected.

  Following the analysis in the usual perturbative expansion \cite{Zak,Dya},
one solution for the occupation numbers $n_{m}$ is given by,
\bea
n_{m} =  \frac{T}{\mu + \varepsilon_{m}} \;\; ,    \label{inteq}
\eea
where $T$ and $\mu$ are constants; this is the large $T$ limit of the
usual thermodynamic distribution $( \exp[(\varepsilon_{k}+\mu)/T] - 1)^{-1}$
\cite{FW} for noninteracting bosons.
The difference in our case is simply
that the effective energy $\varepsilon_{m}$ enters rather than $\omega_{m}$.
Given the precise form of this effective energy,
we see that eqn. (\ref{inteq}) is in fact a self-contained integral
equation for $n_{m}$, but we will not analyze it further here.

   While the preceeding analysis can equally well be carried out in terms
of discrete momenta and sums (which we have done), and continuous
variables with integrals, the examination of the Kolmogorov solutions
requires the later setting. The only care in going over to integral
expressions is in correctly treating $\delta$-function factors. It
is generally the case that the perturbative expansion we have considered
will contain a factor of $\delta(0)$ which must be factored out and
discarded, and it is therefore convenient now to take the
$T_{k_{1}k_{2}k_{3}k_{4}}$ coefficient without the momentum
conserving $\delta$-function, $\delta^{(d)}(k_{1}+k_{2}-k_{3}-k)$,
we previously included so that the integral equivalent of
eqn. (\ref{newnd}) only has a single momentum conserving $\delta$-function.

    One way to establish the Kolmogorov solutions is to follow the
presentation in \cite{Zak,Dya} substituting the effective energy
$\varepsilon_{m}$ for the free field energy $\omega_{m}$. This entails
a number of assumptions. We will assume that the theory has
rotational symmetry which implies, in particular, that the mode number
$n_{k}=n(k)$ and the effective energy $\varepsilon_{k}=\varepsilon(k)$
only depend on the magnitude of the vector $k$. We further
suppose that the scaling properties,
$\varepsilon(\lambda\, k) = \lambda^{\alpha}\, \varepsilon(k)$ and
$T^{\prime}(\lambda\,k_{1},\lambda\,k_{2},
\lambda\,k_{3},\lambda\,k_{4})
= \lambda^{\beta}\; T^{\prime}(k_{1},k_{2},k_{3},k_{4})$ are satisfied,
and that the functional relation $\varepsilon(k)$
is invertible. Without loss of generality, we can take $\varepsilon(0)=0$
by adjusting, say, the $\omega(0)$ coefficient if necessary. Given
these assumptions, we will show that a solution exists of the form
$n(k)\equiv n(\varepsilon) = \varepsilon^{-x}$ (repeated use
of the symbol $n$ for two different functions should cause no confusion;
we will always regard
the mode numbers as functions of the effective energy in the following).

Let us begin by integrating over angles in (\ref{newnd});
the volume element is $d^{d}k_{i} = k_{i}^{d-1}\, dk_{i}\, d\Omega_{i}$
and it should be clear from context whether we use the symbol $k_{i}$
to denote a vector or its magnitude. Using the assumption that we
can invert the relationship $\varepsilon(k)$, we can change variables
in the remaining integrals from $k_{i}$ to $\varepsilon_{i}$.
It is convenient to first define,
\bea
U(\varepsilon_{1},\varepsilon_{2},
  \varepsilon_{3},\varepsilon) &=&
(k_{1}\,k_{2}\,k_{3}\,k)^{d-1} \; |
\frac{d\varepsilon_{1}}{dk_{1}}
\frac{d\varepsilon_{2}}{dk_{2}}
\frac{d\varepsilon_{3}}{dk_{3}}
\frac{d\varepsilon}{dk}  |^{-1}\; \\
& & \int\; |T^{\prime}_{k_{1}k_{2}k_{3}k}|^{2}\;
\; \delta^{(d)}(
k_{1}+k_{2}-k_{3}-k)\; d\Omega_{1}\,d\Omega_{2}\,d\Omega_{3} \;\; , \non
\eea
where we are integrating over three sets of angular variables. Let us
note that
$U$ shares the same symmetry properties as the
coefficient $T^{\prime}$ under permutations of its arguments.
Our task now is to find solutions to,
\bea
0 &=& \int_{0}^{\infty}
d\varepsilon_{1}\,d\varepsilon_{2}\,d\varepsilon_{3}\;
U(\varepsilon_{1},\varepsilon_{2},
  \varepsilon_{3},\varepsilon) \;
  \delta(\varepsilon_{1}+\varepsilon_{2}-
  \varepsilon_{3}-\varepsilon) \label{eee}\\
  & & \;\;\;\;\; n_{k_{1}}\, n_{k_{2}} \, n_{k_{3}} \, n_{k}
\; \{ \, \frac{1}{n_{k_{3}}} + \frac{1}{n_{k}} -
\frac{1}{n_{k_{1}}} - \frac{1}{n_{k_{2}}}  \; \}  \;\;. \non
\eea

It is straightforward to work out the scaling properties of $U$ which
are implied by the assumptions. By simply scaling each
of the momenta on both sides of the defining equation, one quickly
finds the relation,
\bea
U(\lambda\,\varepsilon_{1},\lambda\,\varepsilon_{2},
  \lambda\,\varepsilon_{3},\lambda\,\varepsilon) &=&
\lambda^{\gamma}\;
U(\varepsilon_{1},\varepsilon_{2},
  \varepsilon_{3},\varepsilon) \\
\gamma &=& \frac{3\,d + 2\,\beta}{\alpha} - 4  \;\; . \non
\eea

Returning to the analysis of eqn. (\ref{eee}), one easily uses the
$\delta$-function to carry out the $\varepsilon_{3}$ integral; what
remains is a region $D$ of the
$(\varepsilon_{1},\varepsilon_{2})$-plane.
This region is not the entire first quadrant since we must satisfy
the condition,
\bea
\varepsilon_{3} = \varepsilon_{1} + \varepsilon_{2} - \varepsilon
\ge 0 \;\; .
\eea
Divide $D$ into four sectors as follows,
\bea
D_{1} &=& \{ (\varepsilon_{1},\varepsilon_{2}) \in D \;\; | \;\;
\varepsilon_{1} < \varepsilon, \;\;\; \varepsilon_{2} < \varepsilon \}
\\
D_{2} &=& \{ (\varepsilon_{1},\varepsilon_{2}) \in D \;\; | \;\;
\varepsilon_{1} > \varepsilon, \;\;\; \varepsilon_{2} > \varepsilon \}
\non \\
D_{3} &=& \{ (\varepsilon_{1},\varepsilon_{2}) \in D \;\; | \;\;
\varepsilon_{1} < \varepsilon, \;\;\; \varepsilon_{2} > \varepsilon \}
\non \\
D_{4} &=& \{ (\varepsilon_{1},\varepsilon_{2}) \in D \;\; | \;\;
\varepsilon_{1} > \varepsilon, \;\;\; \varepsilon_{2} < \varepsilon \}
\;\; , \non
\eea
and perform the Zakharov transformations \cite{Dya,VEZ} to map $D_{2}$,
$D_{3}$, and $D_{4}$ onto $D_{1}$. Those transformations respectively
take the form,
\bea
&D_{2}:&\;\;\;
  \varepsilon_{1} = \frac{\varepsilon \varepsilon_{1}^{\prime}}{
  \varepsilon_{1}^{\prime} + \varepsilon_{2}^{\prime} - \varepsilon}
  \;\;\;\;\;
\varepsilon_{2} = \frac{\varepsilon \varepsilon_{2}^{\prime}}{
  \varepsilon_{1}^{\prime} + \varepsilon_{2}^{\prime} - \varepsilon}
  \;\; , \\
&D_{3}:&\;\;\;
\varepsilon_{1} = \frac{\varepsilon (\varepsilon_{1}^{\prime}
+ \varepsilon_{2}^{\prime} - \varepsilon)}{\varepsilon_{2}^{\prime}}
  \;\;\;\;\;
  \varepsilon_{2} = \frac{\varepsilon^{2}}{
  \varepsilon_{2}^{\prime}} \;\; , \non \\
&D_{4}:&\;\;\;
\varepsilon_{1} = \frac{\varepsilon^{2}}{
  \varepsilon_{1}^{\prime}}
  \;\;\;\;\;
  \varepsilon_{2} = \frac{\varepsilon (\varepsilon_{1}^{\prime}
+ \varepsilon_{2}^{\prime} - \varepsilon)}{\varepsilon_{1}^{\prime}}
  \;\; . \non
\eea

Using these transformations, and the ansatz $n(\varepsilon) =
C\, \varepsilon^{-x}$, one finds that eqn. (\ref{eee}) becomes,
\bea
0 &=& \int_{D_{1}}
d\varepsilon_{1}\,d\varepsilon_{2}\;
U(\varepsilon_{1},\varepsilon_{2},
  \varepsilon_{1}+\varepsilon_{2}-\varepsilon,\varepsilon) \;
[ \varepsilon_{1}\, \varepsilon_{2}\,
(\varepsilon_{1}+\varepsilon_{2}-\varepsilon)\,\varepsilon ]^{-x}\\
& &\{ \, (\varepsilon_{1}+\varepsilon_{2}-\varepsilon)^{x} +
\varepsilon^{x} - \varepsilon_{1}^{x} - \varepsilon_{2}^{x}\, \}  \non \\
& &\{ \, 1 + \Big(\frac{\varepsilon_{1}+\varepsilon_{2}
  -\varepsilon}{\varepsilon}\Big)^{y} - \Big(\frac{\varepsilon_{2}}{
  \varepsilon}\Big)^{y} - \Big(\frac{\varepsilon_{1}}{\varepsilon}
\Big)^{y} \}
\;\; , \non
\eea
with $ y = 3x - \gamma - 3 $. We can satisfy this relation at the
four points $x=0$, $x=1$, $y=0$, or $y=1$. The first two cases are simply
limits of the thermodynamic distribution previously considered, while
the last two possibilities correspond to the Kolmogorov solutions,
\bea
n^{(1)}(\varepsilon) = C_{1}\, \varepsilon^{-(\gamma+3)/3} \;\;\; ,
\;\;\;
n^{(2)}(\varepsilon) = C_{2}\, \varepsilon^{-(\gamma+4)/3}
\;\; .
\eea
Although the form of these solutions is the same as in the usual
perturbative analysis, they differ fundamentally in that they
are in fact integral equations for $n(\varepsilon)$, owing
to eqn. (\ref{eps}); we will not pursue an analysis of them here.

\section{Concluding Remarks}

In this paper we have established a different perturbative expansion
which leads to a quantum kinetic equation, similar to the usual one
obtained by expanding about a quadratic Hamiltonian. By placing the
diagonal quartic terms into the unperturbed Hamiltonian, we have
generalized that equation and identified the effective state energy
which essentially takes the place of the free field energy in those earlier
treatments. The framework we establish applies as well to the usual
perturbative situation and
can be considered as an alternative to other derivations.
We have also highlighted some subtleties that arise there in the
leading order terms cubic in the mode numbers and
have offered an argument why one might expect those additional
corrections to be small.

{\Large\bf Acknowledgements}\\
We would like to thank A. Newell for many discussions on weak turbulence.
The work of S.S. is partly supported by Forbairt SC/94/218.

\end{document}